\documentclass[11pt]{article} %
\newcommand{\mcolor}{black}        
\usepackage{graphicx}
\usepackage{fancyhdr}
\usepackage{makeidx}
\usepackage{amssymb,amsmath}
\usepackage{physics}
\usepackage{upquote}
\usepackage{float}
\usepackage{xcolor}
\usepackage[utf8]{inputenc}
\usepackage[nottoc]{tocbibind}
\usepackage{setspace}
\usepackage{enumerate}
\usepackage{textcomp}
\usepackage{gensymb}
\usepackage{xr}
\usepackage{comment}
\usepackage{lineno}
\usepackage{pdfpages}
\usepackage{mathtools}
\usepackage[normalem]{ulem}
\usepackage[export]{adjustbox}
\usepackage[hidelinks,colorlinks,allcolors=.,urlcolor=blue,citecolor=blue]{hyperref}
\usepackage[sorting=none,doi=false,url=false,citestyle=numeric-comp,date=year]{biblatex}
\AtEveryBibitem
{
  \clearfield{note}
  \clearfield{issn}
  \clearfield{eprint}
}
\renewbibmacro{in:}{}
\DeclareFieldFormat{pages}{\mkfirstpage[{\mkpageprefix[bookpagination]}]{#1}}
\newbibmacro{string+doi}[1]{%
  \iffieldundef{doi}{#1}{\href{http://dx.doi.org/\thefield{doi}}{#1}}}
\DeclareFieldFormat{title}{\usebibmacro{string+doi}{\mkbibemph{#1}}}
\DeclareFieldFormat[article]{title}{\usebibmacro{string+doi}{\mkbibquote{#1}}}

\newcommand{\vs}{\vspace}
\newcommand{\hs}{\hspace}

\newcommand{\degreem}{\ensuremath{\degree}}

\newcommand{\mum}{\ensuremath{\mu}}
\newcommand{\rhom}{\ensuremath{\rho}}

\newcommand{\epsilonm}{\ensuremath{\epsilon}}

\newcommand{\lambdam}{\ensuremath{\lambda}}

\newcommand{\nablam}{\ensuremath{\nabla}}

\newcommand{\pmm}{\ensuremath{\pm}}
\newcommand{\simm}{\ensuremath{\sim}}

\newcommand{\phim}{\ensuremath{\phi}}

\newcommand{\psim}{\ensuremath{\psi}}
\newcommand{\Psim}{\ensuremath{\Psi}}

\newcommand{\hslashm}{\ensuremath{\hslash}}

\newcommand\braketma[2]{\ensuremath{\braket{#1}{#2}}}
\newcommand\braketmb[3]{\ensuremath{\langle{#1}\lvert{#2}\lvert{#3}\rangle}}
\ExplSyntaxOn
\newcommand\braketm[3]
  { \tl_if_empty:nTF {#2} { \tl_if_empty:nTF {#3}{\braketma{#1}{#1}}{\braketma{#1}{#3}} } { \braketmb{#1}{#2}{#3} } }
\ExplSyntaxOff

\newcommand{\primem}[1]{\ensuremath{#1^\prime}}

\newcommand{\tsup}{\textsuperscript}
\newcommand{\tsub}{\textsubscript}

\begin{document}
\title{Observation of moir\'e trapped biexciton through sub-diffraction-limit probing using hetero-bilayer on nanopillar}
\author{Mayank Chhaperwal$^{1,||}$, Suman Chatterjee$^{1,2,||}$, Suchithra Puliyassery$^{1}$,\\ Jyothsna Konkada Manattayil$^{1}$, Rabindra Biswas$^{1,3}$, Patrick Hays$^{4}$, Seth Ariel Tongay$^{4}$, \\Varun Raghunathan$^1$, and Kausik Majumdar$^{1*}$\\
	$^1$Department of Electrical Communication Engineering, \\Indian Institute of Science, Bangalore 560012, India\\
    $^2$Currently at Quantum Matter Institute, University of British Columbia,\\ Vancouver, British Columbia V6T 1Z4, Canada\\
    $^3$Currently at Department of Physics, Emory University, Atlanta, USA\\
    $^4$Materials Science and Engineering, School for Engineering of Matter, \\Transport and Energy, Arizona State University, Tempe, Arizona 85287, United States\\ 
	$^{||}$These authors contributed equally\\
	$^*$Corresponding author, email: kausikm@iisc.ac.in}
\date{}
\maketitle
\begin{abstract}
The ability to tune the degree of interaction among particles at the nanoscale is highly intriguing. The spectroscopic signature of such interaction is often subtle and requires special probes to observe. To this end, inter-layer excitons trapped in the periodic potential wells of a moir\'e superlattice offer rich interaction physics, specifically due to the presence of both attractive and repulsive components in the interaction. Here we show that the Coulomb force between two inter-layer excitons switches from repulsive to attractive when the length scale reduces from inter-moir\'e-pocket to intra-moir\'e-pocket in a WS$_2$/WSe$_2$ hetero-bilayer - thanks to the complex competition between direct and exchange interaction. The finding is a departure from the usual notion of repelling inter-layer excitons due to layer polarization. This manifests as the simultaneous observation of an anomalous superlinear power-law of moir\'e exciton and a stabilization of moir\'e trapped biexciton. The experimental observation is facilitated by placing the hetero-bilayer on a polymer-nanopillar/gold-film stack which significantly reduces the inhomogeneous spectral broadening by selectively probing a smaller ensemble of moir\'e pockets compared with a flat sample. This creates an interesting platform to explore interaction among moir\'e trapped excitons and higher order quasiparticles.
\end{abstract}
\newpage
\maketitle
\begin{figure}[H]
	\centering
	\vs{-1in}
	\hs{-0in}
	\includegraphics[width=0.9\paperwidth,center]{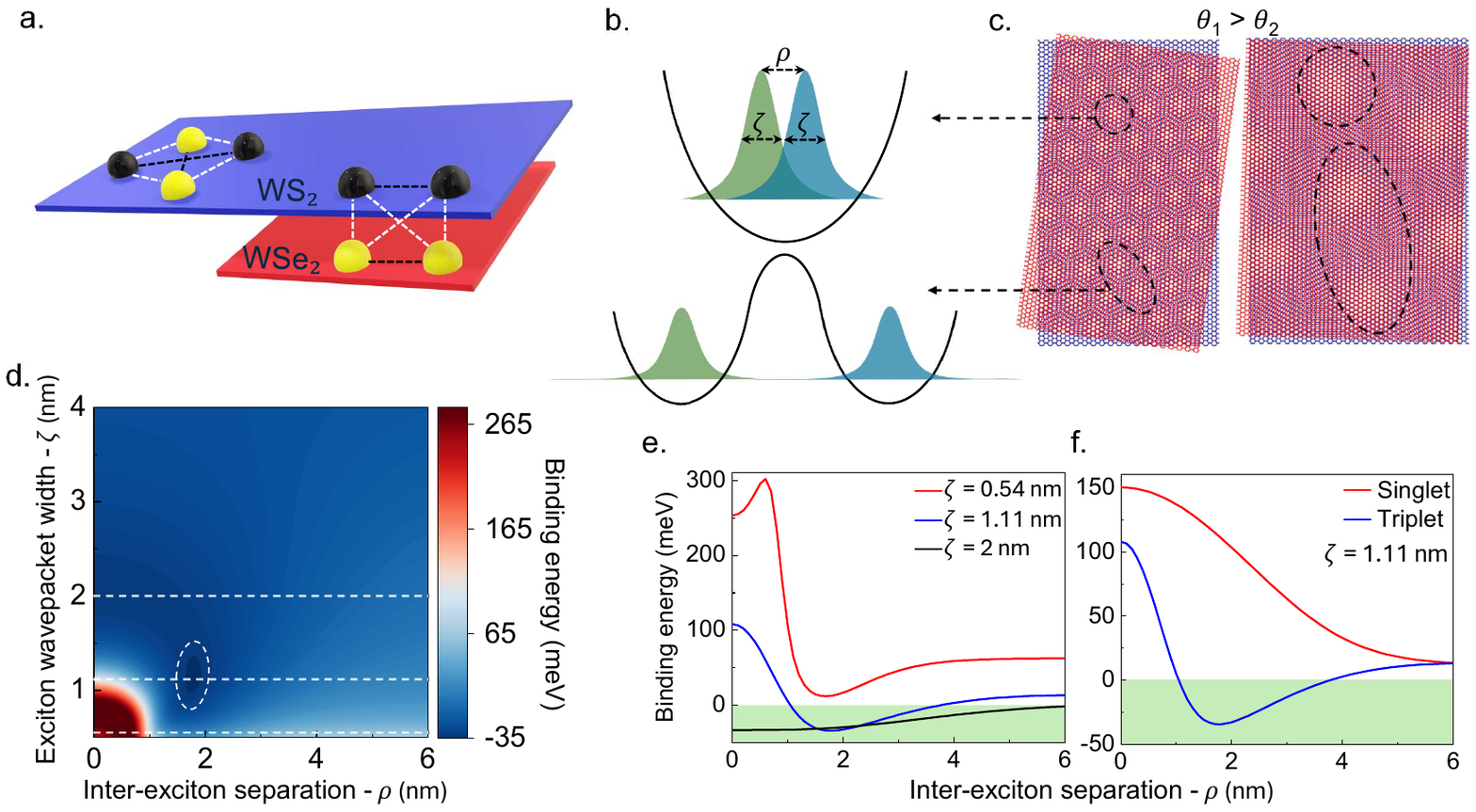}
	\vspace{-0.8in}
	  \caption{\textbf{Stabilization of moir\'e biexcitons:} (a) Schematic diagram showing electron (black spheres) and hole (yellow spheres) spatial arrangement for intra and inter-layer biexcitons in monolayer WS$_2$ and WS$_2$/WSe$_2$ heterojunction, respectively. In the latter case, electrons and holes are layer-polarized, which reduces the degrees of freedom for reconstruction. Black and yellow dashed lines indicate repulsion and attraction, respectively. (b) Schematic representation of two exciton wave packets (separation of $\rho$ and width of $\zeta$) trapped in the same  (top panel) and neighboring (bottom panel) moir\'e well. (c) Moir\'e sample with a higher (left panel) and lower (right panel) twist angle, indicating the effect on moir\'e well size and inter-well separation. (d) A color plot showing the calculated binding energy of the triplet state of the biexciton inside a moir\'e well as a function of $\rho$ and $\zeta$. The white dashed circle indicates the most stable configuration having negative binding energy. (e) Line cuts along the dashed lines in (d). The green shaded region indicates stabilized biexciton regime. (f) Binding energy for singlet and triplet configuration for $\zeta = 1.11$ nm.}\label{fig:model}
\end{figure}
\begin{figure}[H]
	\centering
	\vs{0in}
	\hs{-0in}
    \includegraphics[width=0.9\paperwidth,center]{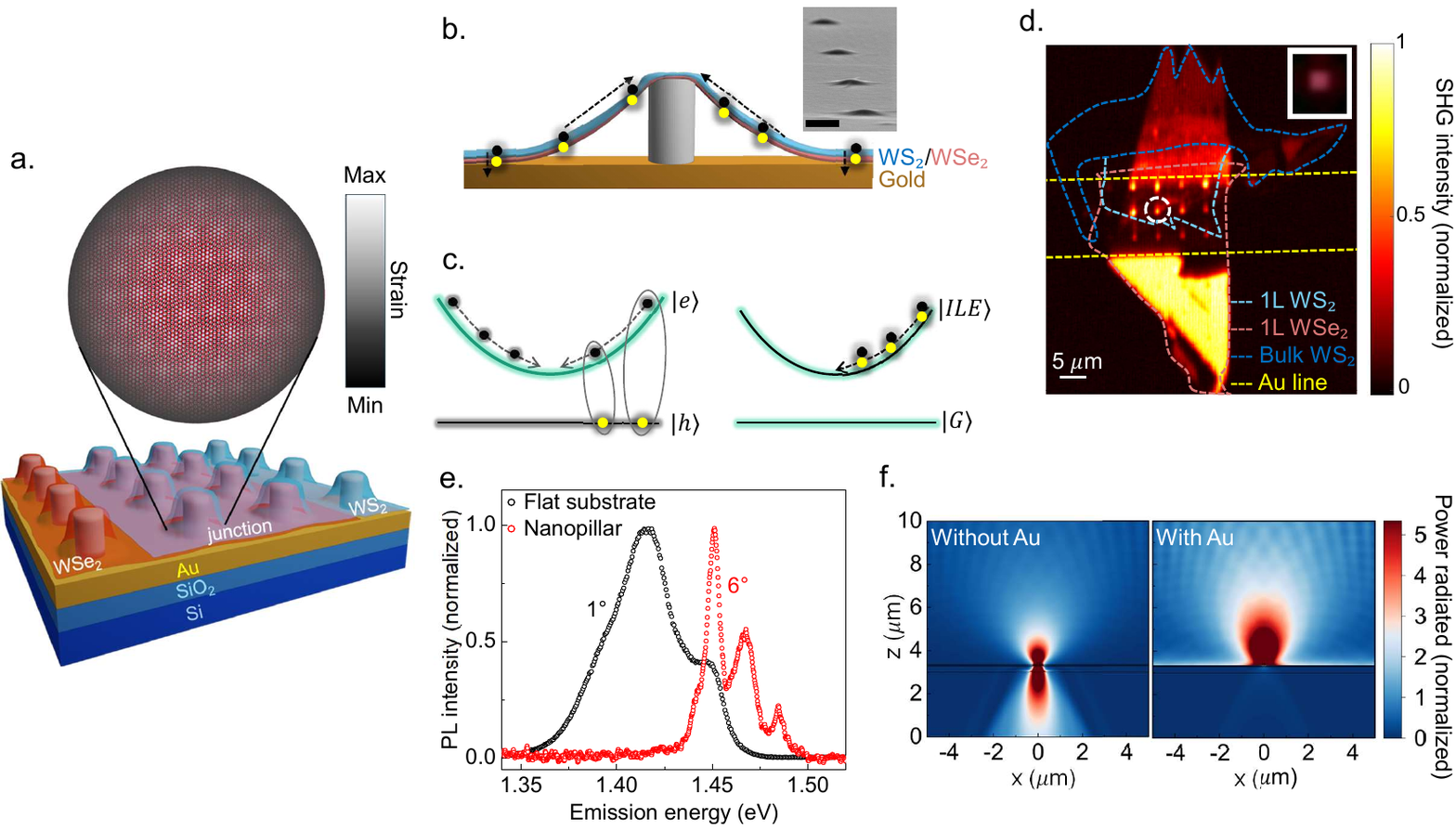}
	\vspace{0in}
	  \caption{\textbf{Strained hetero-bilayer on nanopillar:} (a) Schematic representation of the strained WS$_2$/WSe$_2$ heterojunction on a periodic array of pillars with a gold film underneath. The zoomed-in view with a color bar schematically shows that the highest strain is applied on the pillar head (top panel). (b) Schematic cross-section of the structure. Arrows indicate strain-induced funneling of excitons towards the center of the pillar. On the flat portion, the ILE quenches due to nonradiative transfer to gold (downward arrow). The inset shows an SEM image of the hetero-bilayer on pillars. Scale bar: 500 nm. (c) Funneling (dragging) of electrons (holes) due to the formation of a bound ILE state. The conduction band bending due to strain is significant compared to the valence band. Left panel and right panel show the one-particle and two-particle pictures, respectively. (d) SHG map showing bright emission from pillar heads. Different layers are marked accordingly. Inset: PL map around the marked pillar indicating bright emission from the pillar surrounded by dark region. (e) PL emission spectra from flat (black symbols, twist angle $\sim 1\degree$) and pillar (red symbols, twist angle $\sim 6\degree$) samples at 5 K. (f) Simulation showing gold line acting as an efficient reflector to modify the ILE dipole radiation pattern.}\label{fig:strain}
\end{figure}
\begin{figure}[H]
	\centering
	\vs{0in}
	\hs{0in}
    \includegraphics[width=0.9\paperwidth,center]{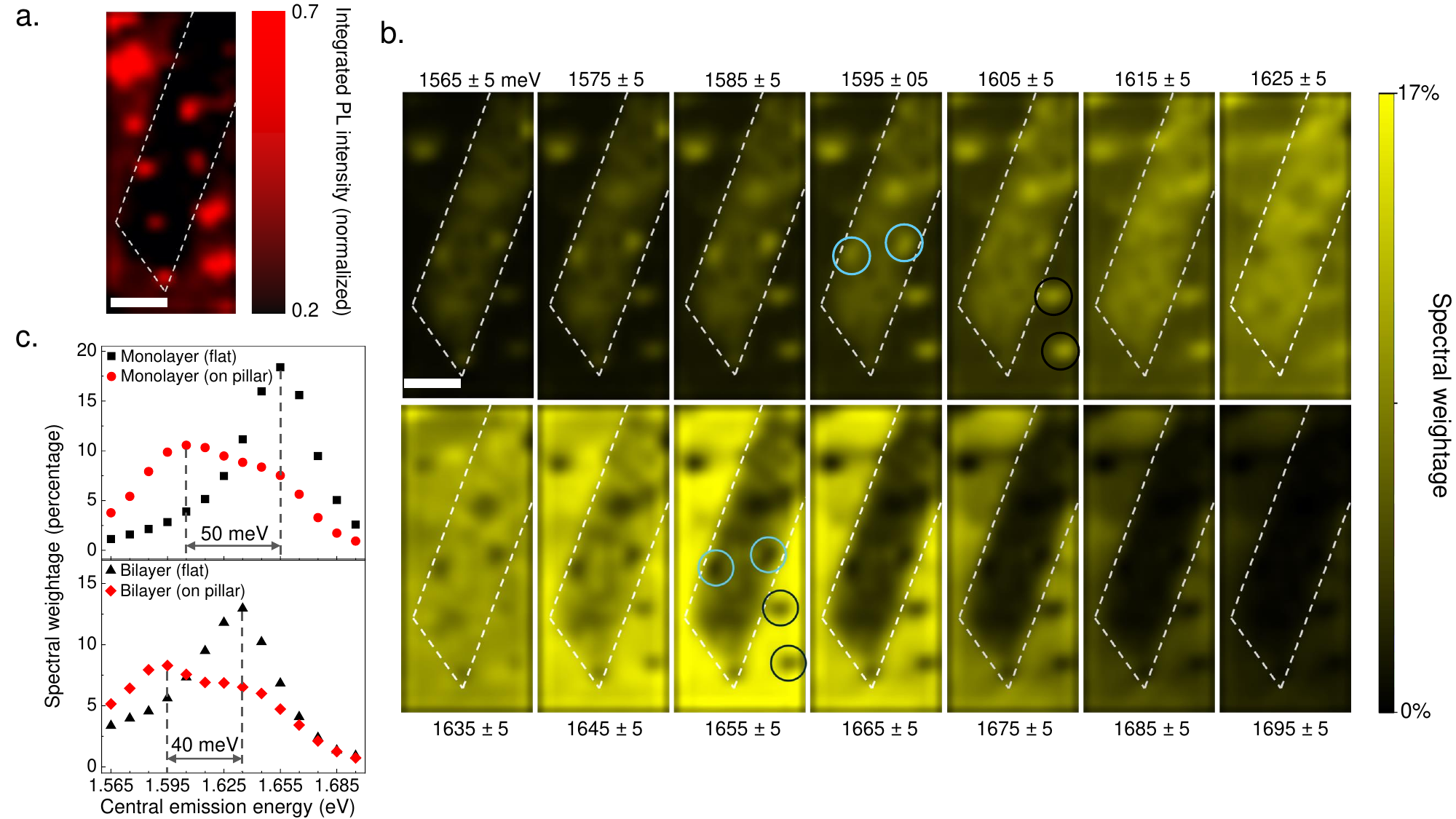}
	\vspace{0in}
	  \caption{\textbf{Quantitative estimation of strain induced bandgap reduction:} (a) Integrated PL intensity map of the sample showing an array of brightly emitting pillars at 300 K. The white dashed line marks the hetero-bilayer region, and the remaining part is monolayer WSe\tsub{2}. Scale bar: 5 \mum{}m (b) Spatial maps show the percentage of photons emitted within the energy ranges specified in each sub-figure (100\% represents the total emission in the entire spectral range at a given spatial point). Black (blue) circles mark some selected pillars in the monolayer (hetero-bilayer) regions. Scale bar: 5 \mum{}m. (c) Spectral weightage for the flat monolayer, flat hetero-bilayer, one of the monolayer covered pillars (black circled), and one of the hetero-bilayer covered pillars (blue circled) is picked from the map in (b) and is plotted as a function of the corresponding  energy range. The strain induced shift in emission energy of monolayer (hetero-bilayer) covered pillars with respect to flat monolayer (hetero-bilayer) portion is mentioned.}\label{fig:map}
\end{figure}
\begin{figure}[H]
	\centering
	\vs{-1in}
	\hs{0in}
    \includegraphics[width=1\paperwidth,center]{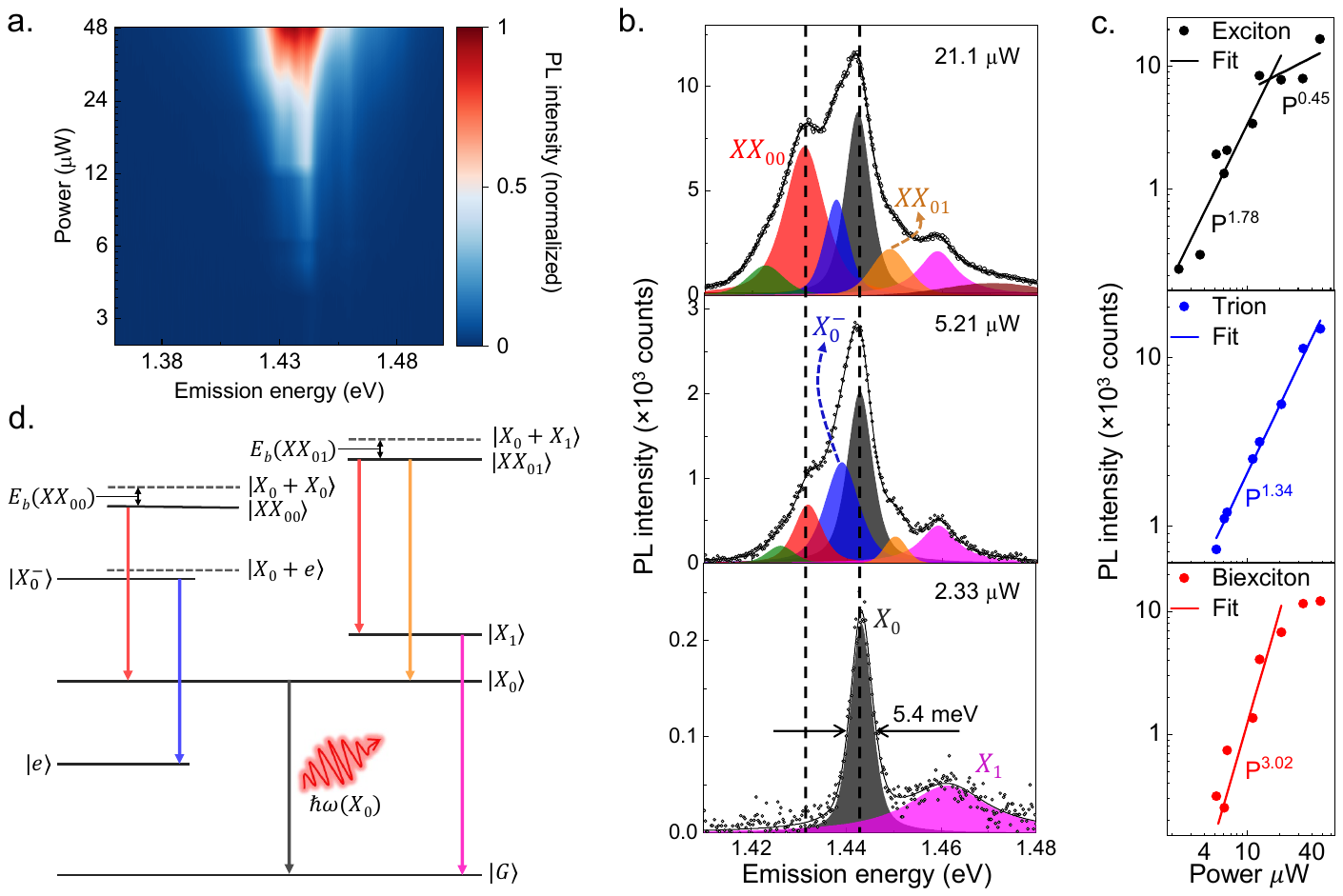}
	\vspace{-0.8in}
	  \caption{\textbf{Features of moir\'e exciton and biexciton emission on pillar:} (a) A color plot showing PL emission as a function of optical power and emission energy at 5 K. (b) Evolution of emission spectra with optical power showing fitted $X_0$ (black), $X_1$ (magenta), $X_0^-$ (blue), $X\!X_{00}$ (red), and $X_{01}$ (orange) peaks.  The black dashed lines are a guide to the eye for $X_0$ and $X\!X_{00}$ peaks. (c) Power law fitting of $X_0$ (top panel), $X_0^-$ (middle panel), and $X\!X_{00}$ (bottom panel) peaks.  (d) Transition diagram showing radiative emission pathways of different states with the same color-coding as in (b).}\label{fig:peaks}
\end{figure}
\begin{figure}[H]
	\centering
	\vs{-0.5in}
	\hs{0in}
    \includegraphics[width=1.3\paperwidth,center]{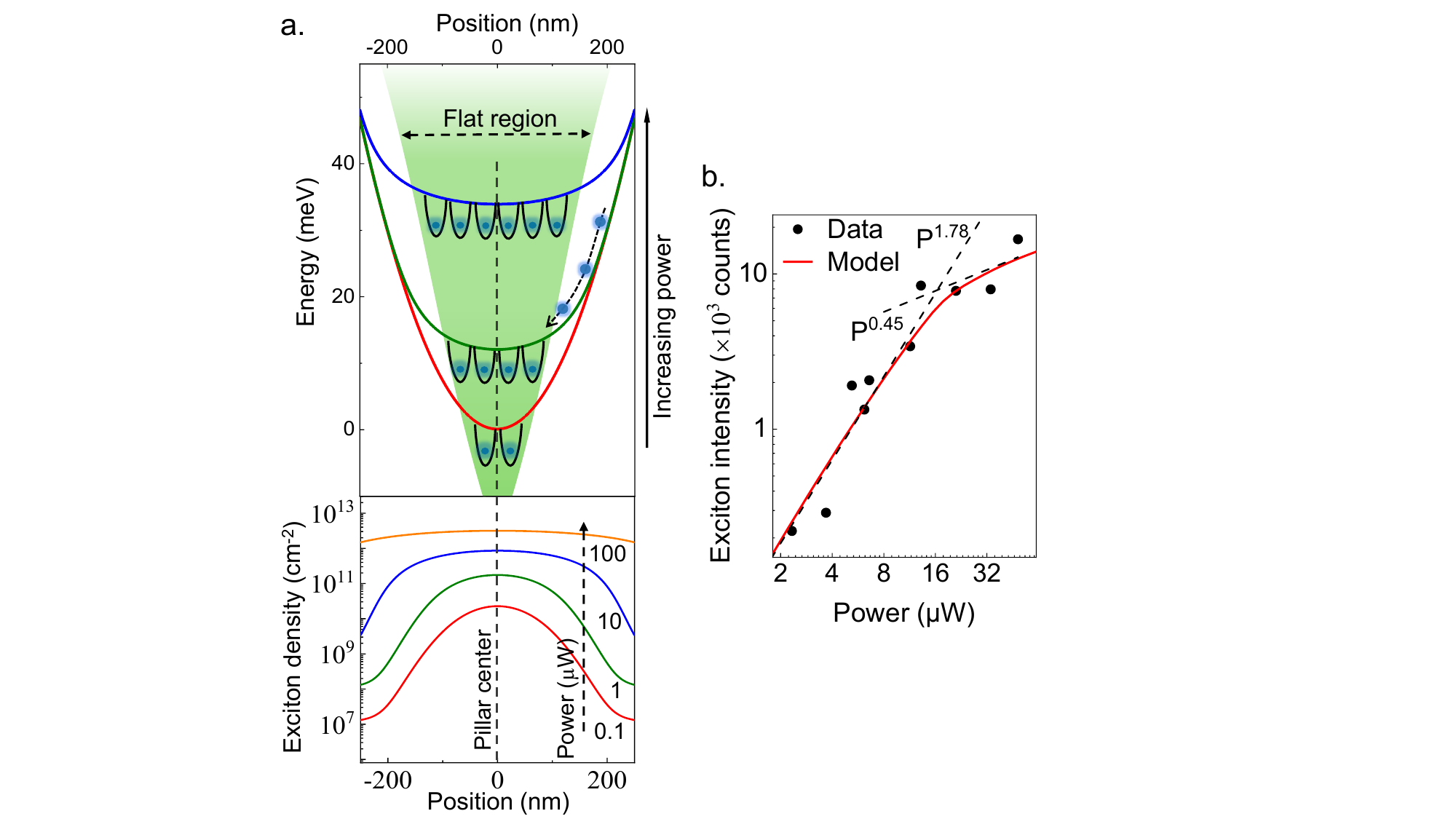}
	\vspace{0in}
	  \caption{\textbf{Optical power dependent modulation of the strain well at pillar site:} (a) Top panel: The dipole-dipole repulsion gives rise to a flattening of the strain well profile, increasing the number of moir\'e wells filled with excitons.  This gives rise to a superlinear power law in \autoref{fig:peaks}c (top panel). Bottom panel: Calculated exciton density as a function of position from the pillar center at different power, suggesting flattening of the profile at higher power. (b) Emission intensity of $X_0$ as a function of power. Symbols indicate measured data, the dashed lines power law fitting, and the red trace shows the fitting from the model.}\label{fig:superlinearity}
\end{figure}

2D Exciton, while being overall charge neutral, is a field-polarizable quasiparticle composed of a negatively charged electron and a positively charged hole that are strongly bound to each other \cite{wangColloquium2018,chernikovExciton2014}. The presence of oppositely charged particles in an exciton makes the inter-exciton interaction highly intriguing due to the complex interplay between attractive and repulsive forces through direct and exchange Coulomb interactions. In a monolayer, two excitons in proximity minimize the total energy by reconstruction (see left part of \autoref{fig:model}a), forming a biexciton that has a negative binding energy \cite{youObservation2015, yeEfficient2018}. In contrast, in the case of inter-layer exciton (ILE) of hetero-bilayer, such a reconstruction is restricted by the strong layer polarization (electron and hole residing in opposite layers, see right part of \autoref{fig:model}a). As a result, inter-exciton interaction often turns out to be repulsive in nature \cite{chatterjeeHarmonic2023,laikhtmanExciton2009,sunEnhanced2022}, making the stabilization of inter-layer biexciton difficult. Such a repulsive interaction has been widely reported through the observation of a blue shift of the ILE emission energy with an increase in the incident optical power \cite{chatterjeeHarmonic2023,sunEnhanced2022}.\\\\
In a moir\'e superlattice, the potential landscape can force two excitons to be trapped inside a single potential well having a diameter on the order of a few nanometers (see \autoref{fig:model}b,c). The dipolar repulsion between the two ILE inside a moir\'e well has been recently shown to create ladder-like higher energy excitonic states \cite{parkDipole2023,bremOptical2024}. This exciton-exciton proximity, coupled with the innate out-of-plane confinement in the ultra-thin layers, results in a competition between the direct and exchange Coulomb interaction. An interesting question to pose here is whether the exchange interaction can overcome the direct repulsive interaction and stabilize the biexciton in a moir\'e well.\\\\
To this end, our variational calculations (discussed later and also in \textbf{Methods}) suggest that such stabilization is indeed possible for the spatially antisymmetric (triplet) biexciton configuration (see \autoref{fig:model}d-f). \autoref{fig:model}d shows the calculated binding energy ($E_b$) of the biexciton as a function of the separation ($\rhom$) between the centers of mass (COM) of the two constituent excitons and the width ($\zeta$) of the exciton wave packets ($\rhom$ and $\zeta$ are defined in \autoref{fig:model}b). $E_b$ becomes negative and exhibits a minimum for small $\rhom$ and $\zeta$ at a length scale similar to the size of the moir\'e well, suggesting the possibility of experimental observation of moir\'e trapped biexciton. On the other hand, the potential barrier between two neighboring moir\'e wells suppresses the wave function overlap (bottom panel of \autoref{fig:model}b), and thus, inter-moir\'e well excitonic interaction remains repulsive due to weak exchange force. Such spatial switching from repulsive to attractive interaction between inter-layer excitons in the moir\'e superlattice opens an avenue for tunable interaction.\\\\
To achieve enhanced interaction among moir\'e excitons in an experimental framework, we need to meet two important criteria:   (a) a larger twist angle that helps in reducing both the size of the moir\'e well and the inter-moir\'e well separation - simultaneously enhancing intra- and inter-moir\'e well Coulomb interaction \cite{zhangTwistangle2020,shabaniDeep2021} (see \autoref{fig:model}c); and (b) a large fill factor that drives such inter-excitonic interaction. Unfortunately, it is challenging to meet both criteria simultaneously in an experimental setup since one must have a large excitation density to achieve a fill factor $>1$ in a sample with dense moir\'e pockets. The large optical power required to achieve such high excitation density causes spectral broadening of the moir\'e trapped ILE emission peaks, smearing out the key spectroscopic signatures of inter-excitonic interactions.  This necessitates a novel moir\'e sample design, as discussed below.\\\\ 
We fabricate twisted WS$_2$/WSe$_2$ hetero-bilayer samples on polymer nanopillar (height: 100 nm, diameter: 150 nm) patterned on a gold film, as shown in \autoref{fig:strain}a (see details of the sample fabrication in \textbf{Methods}). \autoref{fig:strain}b shows a schematic of the cross-section of the structure with a cross-sectional scanning electron micrograph (SEM) in the inset. We keep a relatively large twist angle ($>$5\degreem{}) between the two monolayers, facilitating intra- and inter-moir\'e well Coulomb interaction.\\\\
The pillar structure has several advantages over flat samples: The local strain produced by the nanopillar reduces the bandgap at the top of the pillar, allowing an optically-generated ILE to funnel to the pillar site \cite{Johari2012,Shen2016a,Brooks2018} (\autoref{fig:strain}b,c). This helps us achieve a large fill factor locally at the top of the nanopillar without the need for applying a large optical power.\\\\
The underneath gold film quenches the exciton emission away from the pillar site due to non-radiative exciton transfer to gold, thus allowing us to selectively collect the photons from the top of the pillar \cite{chhaperwalSimultaneously2024,Chaudhary2020}. \autoref{fig:strain}d shows the spatial intensity map of second harmonic generation (SHG) and exciton photoluminescence (PL) signal (in the inset, around the marked pillar) of a representative sample with a twist angle of \simm{}6\degreem{} (confirmed by polarization-resolved SHG polar plot in Supporting Information 1. Both maps indicate that the measured intensity is the maximum on the pillars and is significantly suppressed on gold.\\\\
\textcolor{\mcolor}{The enhanced SHG signal on the pillar with respect to the flat part on gold arises due to a combination of (1) the incident laser's electric field having an antinode near the gold surface and (2) strain on top of the nanopillar \cite{guanStraininduced2025,puriSubstrate2024,changEnhancement2024}. More details are discussed in Supporting Information 2. Both far-field radiation pattern of the SHG signal and the angular distribution of the SHG polarization suggest that the strain on the pillar does not introduce any anisotropy (see Supporting Information 3 for details).}\\\\
\textcolor{\mcolor}{In the case of PL, the enhancement on the nanopillar compared to the flat regions of monolayers touching gold is due to exciton funneling to the pillar site and quenching of excitons away from the pillar. Quenching of excitons away from the pillar is due to multiple reasons, such as, (1) the non-radiative loss of excitons to gold from the flat regions \cite{Chaudhary2020,chhaperwalSimultaneously2024}, (2) excitation laser field having an antinode on the gold surface (see Supporting Information 2), and (3) reduced far-field component of the emission (see Supporting Information 4).} This allows us to collect ILE emission from a much smaller ensemble of moir\'e wells on the pillar compared to a flat sample, the collection area in the latter case being limited by the diffraction-limited laser spot. This dramatically reduces the inhomogeneous broadening of the ILE emission peaks, helping us resolve interaction-driven, closely spaced peaks. \textcolor{\mcolor}{\autoref{fig:strain}e compares the typical emission spectra from a pillar sample and a hBN-capped flat sample. ILE peaks from the pillar sample exhibit a full-width at half-maximum (fwhm) of $\sim 6$ meV - which is a $5$-fold reduction compared with flat samples. See Supporting Information 5 for emission spectrum from another pillar sample with fwhm of 4.9 meV, while the corresponding flat portion on gold of the same sample shows a weak emission peak with fwhm of $\sim 34$ meV.}\\\\
Further, the gold film also acts as a back reflector, reducing the loss of photons emitted in the downward direction (see \autoref{fig:strain}f for emission pattern). Note that the pillars are made of photoresist polymer with a low dielectric constant of 2.62. Further, we avoid capping the sample with hBN to reduce dielectric screening. This low dielectric environment maintains a high degree of Coulomb interaction among the excitons.\\\\
\textcolor{\mcolor}{As reported previously \cite{Niehues2018,Feng2012,Brooks2018,Johari2012}, the conduction band of WS\tsub{2} is tunable with strain, whereas the valence band of WSe\tsub{2} does not change appreciably (left panel of \autoref{fig:strain}c). The strain thus effectively reduces the local band gap and, consequently, the ILE emission energy, as shown in the right panel of \autoref{fig:strain}c. In addition, strain induced selective lowering of the conduction band gives rise to an inward electron flux towards the top of the pillar \cite{Lee2022b,Harats2020}, which leads to a possibility of electron doping at the pillar site.}\\\\
In the rest of the paper, we focus on a sample (with data from additional samples discussed in Supporting Information 6 and Supporting Information 8) whose integrated PL emission map at 300 K is shown in \autoref{fig:map}. The dashed line demarcates the hetero-bilayer region from the WSe\tsub2 monolayer. In both regions, an array of brightly emitting nanopillars is visible, surrounded by a dark background, as expected from the earlier discussion. To estimate the strain-induced bandgap reduction quantitatively,  \autoref{fig:map}b presents spatial maps of the PL intensity, integrated over different energy ranges as indicated. Each spatial point is normalized by the total spectral emission at that point. Data for monolayer, hetero-bilayer, one of the monolayer covered pillars (black circles in \autoref{fig:map}b), and one of the hetero-bilayer covered pillars (blue circles in \autoref{fig:map}b) are picked from the map and plotted in \autoref{fig:map}c for better visualization. The monolayer WSe\tsub{2} portion is the brightest around 1.655 eV, corresponding to the excitonic energy of the monolayer. Note that both the hetero-bilayer portion and the pillars are dark for this range and are brighter in lower energy maps, suggesting a red shift for both of these regions (\autoref{fig:map}b). The flat hetero-bilayer portion becomes the brightest at 1.635 eV (\autoref{fig:map}c) and is thus 20 meV red-shifted compared to the flat monolayer portion. Monolayer-covered pillars exhibit a maximum emission around 1.605 eV, giving a strain-induced redshift of 50 meV compared to the flat monolayer portion. Similarly, the hetero-bilayer-covered pillars show maximum emission at 1.595 eV, giving a strain-induced redshift of 40 meV compared to the flat hetero-bilayer portion.
\\\\
\autoref{fig:peaks} summarizes the key experimental findings of this work. \autoref{fig:peaks}a shows the PL intensity at a pillar site at 5 K in a color plot as a function of the incident power ($P$) of a 532 nm laser and the emission photon energy. The line-cuts at three different $P$ values are shown in \autoref{fig:peaks}b. In \autoref{fig:peaks}c, we show the variation of the intensity ($I$) of different ILE peaks as a function of $P$, and fit them using a power law ($I \propto P^{\gamma}$). The corresponding transitions between states are summarized in \autoref{fig:peaks}d.\\\\\
We look for the peaks that are discernible at the lowest $P$ (bottom panel of \autoref{fig:peaks}b) and assign them as the neutral exciton ($X_0$ and $X_1$) resonances arising from the ground state and first excited state of the moir\'e potential well. Surprisingly, we observe a strong nonlinearity with \textcolor{\mcolor}{$\gamma = 1.78 \pmm{} 0.25$} for $X_0$ (fwhm = 5.4 meV) for nearly a 10-fold change in $P$ (\autoref{fig:peaks}c). Data from another sample exhibiting superlinearity of $X_0$ is given in Supporting Information 6. $X_1$ also exhibits superlinearity with $P$, however, with a reduced degree (\textcolor{\mcolor}{$\gamma = 1.27 \pmm{} 0.11$}, see Supporting Information 7). This is a stark anomaly from conventional neutral exciton power law with $\gamma \approx 1$ for free excitons \cite{youObservation2015,chatterjeeProbing2022,chatterjeeHarmonic2023,Huang2016b} and $\gamma \leq 1$ for trapped excitons \cite{chatterjeeHarmonic2023,chhaperwalSimultaneously2024,heSingle2015,heCascaded2016}. As discussed below, such a superlinear power law is a signature of inter-moir\'e well repulsive dipolar interaction. \\\\
The Coulomb interaction operator between two inter-layer excitons is given by 
\begin{align}
 \hat{V}_{int} &= V(\mathbf{r}_{e1} - \mathbf{r}_{e2}) + V(\mathbf{r}_{h1} - \mathbf{r}_{h2}) -  \primem{V}(\mathbf{r}_{e1} - \mathbf{r}_{h2}) - \primem{V}(\mathbf{r}_{e2} - \mathbf{r}_{h1})
    \label{eq:V}
\end{align}

where,
\begin{equation}
V = \frac{e^2}{\epsilonm{} r} \quad \text{and} \quad \primem{V} = \frac{e^2}{\epsilonm{} \sqrt{r^2 + d^2}}
\label{eq:potential_functions}
\end{equation}
$e$ is the electronic charge, $\epsilon$ is the effective dielectric constant, $d$ is the inter-layer separation, and $r$  is the absolute of the in-plane relative coordinate. The presence of both positive and negative terms gives rise to a complex competition between direct and exchange terms. However, when the excitons are trapped in the neighboring moir\'e wells, the potential barrier in between results in negligible wave function overlap, suppressing the exchange interaction. Thus, in this inter-moir\'e well interaction scenario, the direct Coulomb term dominates, and the inter-layer excitons create a repulsive environment for others trapped in the neighboring moir\'e wells. \\\\
To understand the superlinearity from such repulsive interaction, we calculate the exciton density $n(x,t)$ at a radial distance $x$ from the center of the strain well (see \textbf{Methods} for derivation):
\begin{equation}\label{eq:raten1_gamma}
\frac{\partial n(x,t)}{\partial t} = -\frac{\partial F(x)}{\partial x}+g(x)-\frac{n(x)}{\tau_e}-\lambdam{} n^{2}(x)
\end{equation}
The first term on the right-hand side represents the gradient of the total exciton flux ($F$), which incorporates three key processes: the outward flux of excitons due to (a) concentration gradient and (b) exciton-exciton dipolar repulsion; and their inward drift driven by (c) a strain gradient. The generation rate of excitons, $g(x)$, follows a Gaussian spatial distribution and scales with $P$. The parameter $\tau_e$ is the effective exciton lifetime, encompassing radiative and non-radiative decay mechanisms apart from the Auger process, while \lambdam{} represents the Auger annihilation coefficient. Solving for the steady-state condition ($\displaystyle\frac{\partial n(x,t)}{\partial t} = 0$),  $n(x)$ is determined numerically and is shown in \autoref{fig:superlinearity}a (bottom panel) for different values of $P$. At lower $P$, $n(x)$ peaks at the nanopillar center due to exciton funneling caused by strain. This results in the trapping of excitons within moiré pockets at the lowest energy point of the strain-induced potential (\autoref{fig:superlinearity}a, top panel). As $P$ increases, repulsive dipolar interaction between excitons in adjacent moiré pockets becomes significant, raising the exciton energy with their density \cite{sunExcitonic2022,laikhtmanExciton2009,chatterjeeHarmonic2023}. This elevates the local exciton energy at the center of the nanopillar where the exciton density is maximum, partially offsetting the strain-induced energy lowering. Consequently, the overall energy profile starts to flatten at the center of the nanopillar. The size of this flattened region increases with $P$. The flattening of the energy well gives rise to an increasing number of equipotential moir\'e pockets available for filling (\autoref{fig:superlinearity}a, top panel). This increasing number of emitting moir\'e pockets gives rise to the superlinearity observed in our sample. When the size of the energetically flat region crosses the size of the strain well, the newer moir\'e pockets being filled reside on the metal substrate. These contribute little to the overall emission due to non-radiative charge transfer to gold. This limit, combined with the saturation of the individual moir\'e pocket emission [given by $I=I\tsub{max} \cdot P/(P+P_{sat})$] \cite{heSingle2015,kumarStraininduced2015,tonndorfSinglephoton2015}, leads to the saturation of emission intensity at higher $P$. We overlay the experimental data with emission intensity calculated from the model (\autoref{fig:superlinearity}b) and are able to fit both the initial superlinear increase, followed by the saturating behavior. The observations suggest an optically-driven dynamic tuning of the strain well size and, in turn, effective pillar diameter.\\\\
With an increase in $P$, new emission peaks appear on both lower and higher energy sides of $X_0$. On the lower energy side, we observe a trion ($X_0^-$), about 4.5 meV below the $X_0$ peak - in good agreement with previously reported moir\'e trions \cite{liuSignatures2021,wangMoire2021}. The $X_0^-$ peak can be characterized with a \textcolor{\mcolor}{$\gamma=1.34 \pmm{} 0.05$} (\autoref{fig:peaks}c, middle panel). The appearance of $X_0^-$ with an increase in $P$ is due to the photoinduced doping that results from the additional photoelectrons funneling towards the pillar site due to the strain induced curvature of the conduction band edge (left panel of \autoref{fig:strain}c) \cite{desaiStrainInduced2014,Niehues2018}. Note that the spectral location of $X_0$ does not change appreciably over a wide range of power - an observation that is made in all our samples. This indicates that with an increase in $P$, the photo-electron doping induced red shift \cite{chatterjeeHarmonic2023} likely compensates for the dipolar repulsion driven blue shift. 
\\\\
At higher $P$, we observe an intense peak ($X\!X_{00}$) below $X_0^-$. $X\!X_{00}$ exhibits a power law index of \textcolor{\mcolor}{3.02 \pmm{} 0.48} (\textcolor{\mcolor}{1.68 \pmm{} 0.12}) with respect to $P$ ($X_0$) (\autoref{fig:peaks}c, bottom panel). Such strong superlinearity with respect to $X_0$ indicates the biexcitonic origin of the peak that results from the radiative transition $X\!X_{00} \rightarrow X_0$ (see \autoref{fig:peaks}d). The corresponding emission energy $\hslashm\omega(X\!X_{00})$ being energetically lower than the $X_0$ resonance clearly suggests a negative binding energy ($\sim$11 meV), leading to a stabilization of the biexciton inside a moir\'e well when local fill factor is more than unity. Results from another sample are shown in Supporting Information 8, clearly exhibiting prominent $X_0$, $X_0^-$, and $X\!X_{00}$ emission peaks. Note that such a scenario of biexciton stabilization is distinct from previous works demonstrating two excitons forcefully confined by the moir\'e potential well, which would otherwise be unbounded \cite{parkDipole2023,bremOptical2024}.\\\\
To understand the origin of such intra-moir\'e attractive interaction, we use variational calculations (see \textbf{Methods} for details) to estimate the binding energy of the biexciton. We construct the spatial part of the singlet (triplet) form [indicated by the upper (lower) sign] of the total wave function of the two-exciton state, maintaining symmetry (anti-symmetry) under the exchange of position of either electron or hole:
\begin{align}
\Psim^{\pm}{(\mathbf{r}\tsub{e1}, \mathbf{r}\tsub{h1}, \mathbf{r}\tsub{e2}, \mathbf{r}\tsub{h2})} &\propto \psim{(\mathbf{r}\tsub{e1}, \mathbf{r}\tsub{h1})}\phim{(\mathbf{r}\tsub{e2}, \mathbf{r}\tsub{h2})} \pm \psim{(\mathbf{r}\tsub{e2}, \mathbf{r}\tsub{h1})}\phim{(\mathbf{r}\tsub{e1}, \mathbf{r}\tsub{h2})} \nonumber\\& \pm \psim{(\mathbf{r}\tsub{e1}, \mathbf{r}\tsub{h2})}\phim{(\mathbf{r}\tsub{e2}, \mathbf{r}\tsub{h1})} + \psim{(\mathbf{r}\tsub{e2}, \mathbf{r}\tsub{h2})}\phim{(\mathbf{r}\tsub{e1}, \mathbf{r}\tsub{h1})}
\end{align}
Here $\psim{}$ and $\phim{}$ denote the two individual exciton wave functions (parameterized by inter-exciton separation $\rhom$ and width of exciton wave packet $\zeta$, see \textbf{Methods}) with corresponding electron (hole) coordinates denoted by $\mathbf{r}\tsub{e(h)i}$ for the i\tsup{th} exciton. With $\hat{H}$ being the total Hamiltonian, we obtain the energy of the biexciton state as 
\begin{align}\label{eq:total_energy}
    U^\pm(\rhom,\zeta) &= \braketm{\Psim^\pm{(\mathbf{r}\tsub{e1},\mathbf{r}\tsub{h1},\mathbf{r}\tsub{e2},\mathbf{r}\tsub{h2})}}{\hat{H}}{\Psim^\pm{(\mathbf{r}\tsub{e1},\mathbf{r}\tsub{h1},\mathbf{r}\tsub{e2},\mathbf{r}\tsub{h2})}}
    \nonumber\\&=\int\!d\mathbf{r}\tsub{e1} d\mathbf{r}\tsub{h1} d\mathbf{r}\tsub{e2} d\mathbf{r}\tsub{h2}\ \Psim^{\pm *}{(\mathbf{r}\tsub{e1},\mathbf{r}\tsub{h1},\mathbf{r}\tsub{e2},\mathbf{r}\tsub{h2})} \hat{H}\Psim^\pm{(\mathbf{r}\tsub{e1},\mathbf{r}\tsub{h1},\mathbf{r}\tsub{e2},\mathbf{r}\tsub{h2})}
\end{align}
For large $\rhom{}$, the inter-exciton interaction becomes negligible, and thus $U^\pm\xrightarrow{}$ 2$E_{X_0}$. Therefore, the binding energy of the singlet (upper sign) and triplet (lower sign) biexciton can be estimated for a given $\rhom$ and $\zeta$ as
\begin{equation}\label{eq:BindingEnegry}
E_b^\pm = [\lim_{\rhom{} \to\infty} U^\pm(\rhom,\zeta_0)] - U^\pm(\rhom,\zeta)
\end{equation}
where $\zeta_0$ is the width of a single exciton wave packet, in the absence of any interaction with the other exciton.\\\\
In \autoref{fig:model}d, we plot the triplet biexciton binding energy as a function of the variational parameters $\zeta$ and $\rhom$. Three line-cuts at different $\zeta$ are shown in \autoref{fig:model}e. Large $\zeta$ represents excitons that are not trapped inside a moir\'e pocket. From the repulsive nature of the interaction potential (Equation \ref{eq:V}), we expect the triplet state to have lower energy than the singlet state, as evidenced by the plot in \autoref{fig:model}f. Interestingly, the calculations indicate that the energy minimum of the triplet state occurs for a length scale where two excitons are inside a single moir\'e well in our samples, suggesting that it is possible to achieve a stabilization of the biexciton state with $E_b<0$, thanks to the strong exchange interaction.\\\\
Note that, the calculations suggest that the biexciton formation leads to the pulling of the two constituent excitons close to the center of the moir\'e pocket (although, they do not exactly collapse to the center as the repulsion eventually starts dominating, as depicted in the bottom left corner of \autoref{fig:model}d). This, in turn, creates a strong wave function overlap between $X\!X_{00}$ and $X_0$, creating a high radiative decay rate for the $X\!X_{00} \rightarrow X_0$ transition. On the other hand, as pointed out recently \cite{bremOptical2024}, had the two excitons experienced a net repulsive interaction, they would be pushed towards the opposite edges of the moir\'e pocket. This would reduce the wave function overlap between the initial $X\!X_{00}$ state and the final $X_0$ state during the radiative transition, suppressing $X\!X_{00}$ luminescence. Thus, the experimental observation of strong photoluminescence intensity from the $X\!X_{00}$ peak in our sample is a signature of the attractive intra-moir\'e excitonic interaction.\\\\
\textcolor{\mcolor}{Further, the time-resolved photoluminescence (TRPL) measurement of $X_0$ and $X\!X_{00}$ (see Supporting Information 9) provides additional signature of moir\'e biexciton formation. We observe a non-monoexponential decay for both the species, coupled with a delayed luminescence from the biexciton with respect to the exciton, suggesting biexciton formation from excitons. Both these results are in good agreement with our model that incorporates coupled nonlinear differential equations describing the dynamics of $X_0$ and $X\!X_{00}$ (see Supporting Information 9). In addition, we also observe a fast initial decay ($<100$ ps) for the exciton, unlike flat samples \cite{choiTwist2021, chatterjeeHarmonic2023}, suggesting a fast conversion channel of exciton to biexciton, especially at higher exciton density.} \\\\
At high $P$, more peaks appear (shaded in orange and green in \autoref{fig:peaks}b) in the emission spectrum with superlinear power law - indicative of multi-particle excitonic states. Further investigation is necessary to assign the specific origin of these peaks.  The orange peak around 1.45 eV is particularly interesting, with \textcolor{\mcolor}{$\gamma=1.68 \pmm{} 0.19$} (see Supporting Information 7). We assign this state as the biexciton ($X\!X_{01}$) formed from the excitons at $X_0$ and $X_1$ states. There are two recombination pathways of the $X\!X_{01}$, namely, (a) $X\!X_{01} \rightarrow X_0$ (higher emission energy) and (b) $X\!X_{01} \rightarrow X_1$ (lower emission energy, see \autoref{fig:peaks}d). The former corresponds to the orange shaded peak. A similar separation among the pairs $X_0$ - $X\!X_{00}$ and $X_1$ - $X\!X_{01}$ suggests similar binding energy of $X\!X_{00}$ and $X\!X_{01}$. This also indicates that the photons emitted from the latter transition are energetically close to the $X\!X_{00} \rightarrow X_0$ emission and hence are likely spectrally masked by the bright $X\!X_{00}$ emission. \\\\
To conclude, using a strained hetero-bilayer on a nanopillar, we demonstrate that the exchange force can overcome repulsive interaction between two inter-layer excitons and form a stable triplet biexciton inside a single moir\'e well.  This could lead to possible many-body explorations such as the formation of droplets of moir\'e excitons and interaction among droplets trapped in individual moir\'e pockets.

\section*{Methods}
\subsection*{Variational method to estimate the stability of biexciton}
We start with considering the wave function of the individual moir\'e excitons exhibiting Gaussian wave packet with an s-state like structure \cite{parkDipole2023}. We take two exciton wave packets which are symmetrically positioned around the center of the moir\'e well, and the corresponding wave functions are of the form 
$\psim{} = Ke^{-\frac{(R-\rhom/2)^{2}}{2\zeta^{2}}}e^{-\frac{r}{a_0}}$ and $\phim{} = Ke^{-\frac{(R+\rhom/2)^{2}}{2\zeta^{2}}}e^{-\frac{r}{a_0}}$, where $\mathbf R = \frac{\mathbf r\tsub{e}+\mathbf r\tsub{h}}{2}$ and $\mathbf{r} = \frac{\mathbf r\tsub{e}-\mathbf r\tsub{h}}{2}$, $\zeta$ is the width of the Gaussian wave packet, $a_{0}$ is the Bohr radius of the exciton, $\rhom$ is the separation between the two excitons along the radial direction of the moir\'e well, and $K$ is a normalization constant. \textcolor{\mcolor}{Such a simplified form of the wave function may not necessarily obey the complex symmetry of a moir\'e well after relaxation, especially in the presence of local inhomogeneity. However, this is good enough to elucidate the essential physics required to understand the binding of the excitons within a moir\'e well, particularly at slightly larger twist angles (as is the case in this work).} For single exciton, we assume $a_0=2$ nm, and $\zeta\equiv\zeta_0=2$ nm \cite{parkDipole2023}. $\mathbf r\tsub{e}$ and $\mathbf r\tsub{h}$ are the electron and hole spatial coordinates. We assume that the Bohr radius of individual exciton remains unchanged during exciton-exciton interaction. We construct the spatial part of the singlet (triplet) form [indicated by the upper (lower) sign] of the total wave function of the two-exciton state, maintaining symmetry (anti-symmetry) under the exchange of position of either electron or hole:
\begin{align}
\Psim^\pm{(\mathbf{r}\tsub{e1}, \mathbf{r}\tsub{h1}, \mathbf{r}\tsub{e2}, \mathbf{r}\tsub{h2})} &= K_0[\psim{(\mathbf{r}\tsub{e1}, \mathbf{r}\tsub{h1})}\phim{(\mathbf{r}\tsub{e2}, \mathbf{r}\tsub{h2})} \pm \psim{(\mathbf{r}\tsub{e2}, \mathbf{r}\tsub{h1})}\phim{(\mathbf{r}\tsub{e1}, \mathbf{r}\tsub{h2})} \nonumber\\& \pm \psim{(\mathbf{r}\tsub{e1}, \mathbf{r}\tsub{h2})}\phim{(\mathbf{r}\tsub{e2}, \mathbf{r}\tsub{h1})} + \psim{(\mathbf{r}\tsub{e2}, \mathbf{r}\tsub{h2})}\phim{(\mathbf{r}\tsub{e1}, \mathbf{r}\tsub{h1})}]
\end{align}
where $K_0$ is the normalization constant, and the corresponding electron (hole) coordinates are denoted by $\mathbf{r}\tsub{e(h)i}$ for the i\tsup{th} exciton.\\
With this construction, we ensure that:\\
For exciton exchange ($\mathbf r\tsub{e1} \xleftrightarrow{} \mathbf r\tsub{e2}$ and $\mathbf r\tsub{h1} \xleftrightarrow{} \mathbf r\tsub{h2}$): $\Psim^\pm{} \xrightarrow{} \Psim^\pm{}$
\\For electron exchange ($\mathbf r\tsub{e1} \xleftrightarrow{} \mathbf r\tsub{e2}$): $\Psim^\pm{} \xrightarrow{} \pm\Psim^\pm{}$
\\For hole exchange ($\mathbf r\tsub{h1} \xleftrightarrow{} \mathbf r\tsub{h2}$): $\Psim^\pm{} \xrightarrow{} \pm\Psim^\pm{}$
\\\\
Now, the total Coulomb potential operator is given by:
\begin{align}
 \hat{V} &= V(\mathbf{r}_{e1} - \mathbf{r}_{e2}) + V(\mathbf{r}_{h1} - \mathbf{r}_{h2}) - \primem{V}(\mathbf{r}_{e1} - \mathbf{r}_{h1}) - \primem{V}(\mathbf{r}_{e2} - \mathbf{r}_{h2}) \nonumber\\& - \primem{V}(\mathbf{r}_{e1} - \mathbf{r}_{h2}) - \primem{V}(\mathbf{r}_{e2} - \mathbf{r}_{h1})
    \label{eq:V_total}
\end{align}\\
where,
\begin{equation}
V(\mathbf r) = \frac{e^2}{\epsilonm{} r} \quad \text{and} \quad \primem{V}(\mathbf r) = \frac{e^2}{\epsilonm{} \sqrt{r^2 + d^2}}
\label{eq:potential_functions}
\end{equation}\\
The full Hamiltonian, including the kinetic energy terms for all four particles, is then
\begin{equation}\label{eq:Hamiltonian}
\hat{H} = - \frac{\hslashm{}^2}{2m\tsub{e}}\nablam{^2}_{\mathbf r\tsub{e1}} - \frac{\hslashm{}^2}{2m\tsub{e}}\nablam{^2}_{\mathbf r\tsub{e2}} - \frac{\hslashm{}^2}{2m\tsub{h}}\nablam{^2}_{\mathbf r\tsub{h1}} - \frac{\hslashm{}^2}{2m\tsub{h}}\nablam{^2}_{\mathbf r\tsub{h2}} + \hat{V}
\end{equation}

The energy of the four particle system is then:
\begin{align}\label{eq:total_energy}
    U^\pm(\rhom,\zeta) &= \braketm{\Psim^\pm{(\mathbf{r}\tsub{e1},\mathbf{r}\tsub{h1},\mathbf{r}\tsub{e2},\mathbf{r}\tsub{h2})}}{\hat{H}}{\Psim^\pm{(\mathbf{r}\tsub{e1},\mathbf{r}\tsub{h1},\mathbf{r}\tsub{e2},\mathbf{r}\tsub{h2})}}
    \nonumber\\&=\int\!d\mathbf{r}\tsub{e1} d\mathbf{r}\tsub{h1} d\mathbf{r}\tsub{e2} d\mathbf{r}\tsub{h2}\ \Psim^{\pm *}{(\mathbf{r}\tsub{e1},\mathbf{r}\tsub{h1},\mathbf{r}\tsub{e2},\mathbf{r}\tsub{h2})} \hat{H}\Psim^{\pm}{(\mathbf{r}\tsub{e1},\mathbf{r}\tsub{h1},\mathbf{r}\tsub{e2},\mathbf{r}\tsub{h2})}
\end{align}

For large $\rhom{}$, the inter-exciton interaction becomes negligible, and thus $U^\pm\xrightarrow{}$ 2$E_X$. Therefore, the binding energy of the biexciton can be estimated for a given $\rhom$ and $\zeta$ as
\begin{equation}\label{eq:BindingEnergy}
E_b^\pm = [\lim_{\rhom{} \to\infty} U^\pm(\rhom,\zeta_0)] - U^\pm(\rhom,\zeta)
\end{equation}
We vary $\rhom$ and $\zeta$ as variational parameters. We plot $E_b$ in \autoref{fig:model}d as a function of $\rhom$ and $\zeta$, and look for the minimum of $E_b$. For calculations, we have assumed $d=0.6$ nm and $\epsilon=6.5\epsilon_0$, where $\epsilon_0$ is the permittivity of vacuum. \textcolor{\mcolor}{The stability of the biexciton is analyzed in  Supporting Information 10 for different values of $d$ and $\epsilon$. Further, in Supporting Information 11, we show the role of twist angle (and hence, moir\'e well size) on the stability of the biexciton.}

\subsection*{Numerical solution of combined drift-diffusion equation}
Nanopillars introduce a non-uniform strain on the TMD monolayer, which can be considered parabolic with its maximum in the center of the nanopillar. The band gap reduction can thus be modelled as a parabolic function with its minimum at the center of the nanopillar. The energy of the ILE ($E_{ILE}$) in the presence of such parabolic strain can be written as:
\begin{equation} \label{eq:strain parabolic model}
E_{ILE} = E_g - \frac{1}{2} k x^2
\end{equation}
Here $k$ quantifies the relation between the magnitude of strain and the band gap reduction, $E\tsub{g}$ is the inter-layer bandgap for a flat sample, and $x$ is the radial distance from the center of the pillar. Dipole-dipole repulsion can be taken into account by addition of the term $\alpha_{dd}\,n(x)$ where $\alpha_{dd}$ is the strength of dipole-dipole repulsion and $n(x)$ is the ILE density in the system \cite{sunExcitonic2022,laikhtmanExciton2009,chatterjeeHarmonic2023}. The overall energy of the ILE in the system is then given by:
\begin{equation}\label{eq:total energy}
E_{ILE} = E_g - \frac{1}{2} k x^2 + \alpha_{dd} n(x)
\end{equation}
There will be a drift of ILE due to this non-uniform band gap, which can be modeled as a drift under an induced electric field given by:

\begin{equation}\label{eq:ElectricField}
E_{F}(x) = \frac{-1}{{e}}\frac{\partial E_{ILE}}{\partial x}
\end{equation}

Taking into account the diffusion of ILE to get the total flux ($F$) of excitons at a particular point in the system:
\begin{equation}\label{eq:rate equation}
F = \mu_e n E_{F} - D_e\frac{\partial n}{\partial x} 
\end{equation}
$\mu_e$ and $D_e$ are the drift mobility and diffusion constant for excitons, respectively.

Considering other factors, such as the radiative or nonradiative decay of ILE (overall lifetime $\tau_e$), Auger annihilation (rate constant \lambdam{}), and the generation rate from optical pumping [$g(x)$], the rate of change of ILE density in space and time is thus given by the following rate equation:
\begin{equation}\label{eq:full rate equation}
\frac{\partial n(x)}{\partial t} = -\frac{\partial F}{\partial x} + g(x) - \frac{n(x)}{\tau_e} - \lambdam{} n^2(x)
\end{equation}
Values of $\mu_e$, $D_e$, and \lambdam{} are taken from literature \cite{Cadiz2018b,Dirnberger2021,yuanExciton2015}.

\subsection*{Sample fabrication}
A 30/40 nm thick Ti/Au layer is deposited onto a Si substrate (coated with 285 nm thick thermally grown SiO\tsub{2}) using a sputtering process. A layer of negative tone resist (ma-N-2403 from micro resist technology) is spin-coated on the substrate. The resist is then cured by heating the substrate at 90°C for 2 minutes. A pattern of solid circles is defined on the resist-coated substrate through electron beam lithography, with an accelerating voltage of 30 kV and a 10 µm aperture. The circles in the pattern have a diameter of 150 nm, arranged in an array with a spacing of 5 µm (which is well beyond our laser spot size so that, at a time, only one pillar is excited by the laser during measurement). The electron beam selectively hardens the resist in the areas corresponding to the circular pattern. Following this, the substrate is developed using AZ-726 MIF developer to remove the unexposed resist. To enhance the mechanical durability of the nanopillars, the substrate undergoes a hard-bake process at 150°C for 15 minutes, preparing it for subsequent layer transfer steps.
\\
Both WSe$_2$ and WS$_2$ layers are exfoliated on separate polydimethylsiloxane (PDMS) sheets, and appropriate monolayer flakes are identified under a microscope by optical contrast. We then use micro-manipulators to transfer these flakes one by one onto the patterned nanopillars. The entire structure is then annealed in vacuum (10$^{-6}$ mbar) at 150$^{\circ}$C for 3 hours.
\subsection*{Photoluminescence experiments}
The PL measurements are carried out at 5 K temperature (unless otherwise stated) with a ×50 objective (numerical aperture of 0.5). The excitation source used is a 532 nm laser operated in CW mode. The spot size for the laser is $\approx$ 1.5 $\mu$m. PL spectra are recorded with a spectrometer consisting of a grating with 1800 lines per mm and CCD.
\subsection*{\textcolor{\mcolor}{Far-field SHG intensity pattern imaging}}
\textcolor{\mcolor}{The angular distribution of the SHG intensity from the sample was characterized using a back focal plane (BFP) imaging method. A 4f-lens imaging setup consists of a 180 mm tube lens followed by a 50 mm lens, which is used to project and map the back focal plane of the collection objective (20x/0.75 numerical aperture) onto a Peltier cooled Electron-Multiplying Charge-Coupled Device (EMCCD, iXon Ultra 897) detector plane. An appropriate shortpass (890 nm) and bandpass (520 \pmm{} 20 nm) filter are used to reject residual pump power and minimize background signal.}
\section*{Author contributions}
MC and SC contributed equally to this work.
KM designed the experiment. MC and SC fabricated the samples and performed the measurements. PH and SA provided WSe\tsub2 crystals for exfoliation. MC, SC and KM carried out the data analysis. SEM imaging was performed by SP. JKM, RB and VR performed the SHG measurements. MC and KM carried out the calculations. MC, SC and KM wrote the manuscript with input from all authors. 
\section*{Data availability statement}
Data available on reasonable request from the corresponding author.
\section*{Competing Interests}
The authors declare no competing financial interest.
\section*{Acknowledgments}
M.C. and K.M. acknowledge useful discussions with Varun Srivatsav Kondapally. This work was supported in part by National Quantum Mission, an initiative of the Department of Science and Technology (DST), Government of India, a Core Research Grant from the Science and Engineering Research Board (SERB) under DST, a grant from Indian Institute of Science under IoE, grants from Indian Space Research Organization (ISRO), a grant from DRDO, a grant under SERB TETRA, a grant from I-HUB QTF, IISER Pune, and a seed funding under Quantum Research Park (QuRP) from Karnataka Innovationand Technology Society (KITS), K-Tech, Government of Karnataka. S.A.T. acknowledges direct support from DOE-SC0020653 (excitonic testing of TMD crystals) and NSF CBET 2330110 (environmental testing of TMDs). SAT also acknowledges support from Applied Materials and Lawrence Semiconductors.
\printbibliography

\newpage
\newpage
\newpage
\newpage
\newpage
\newpage
\includepdf[pages=-]{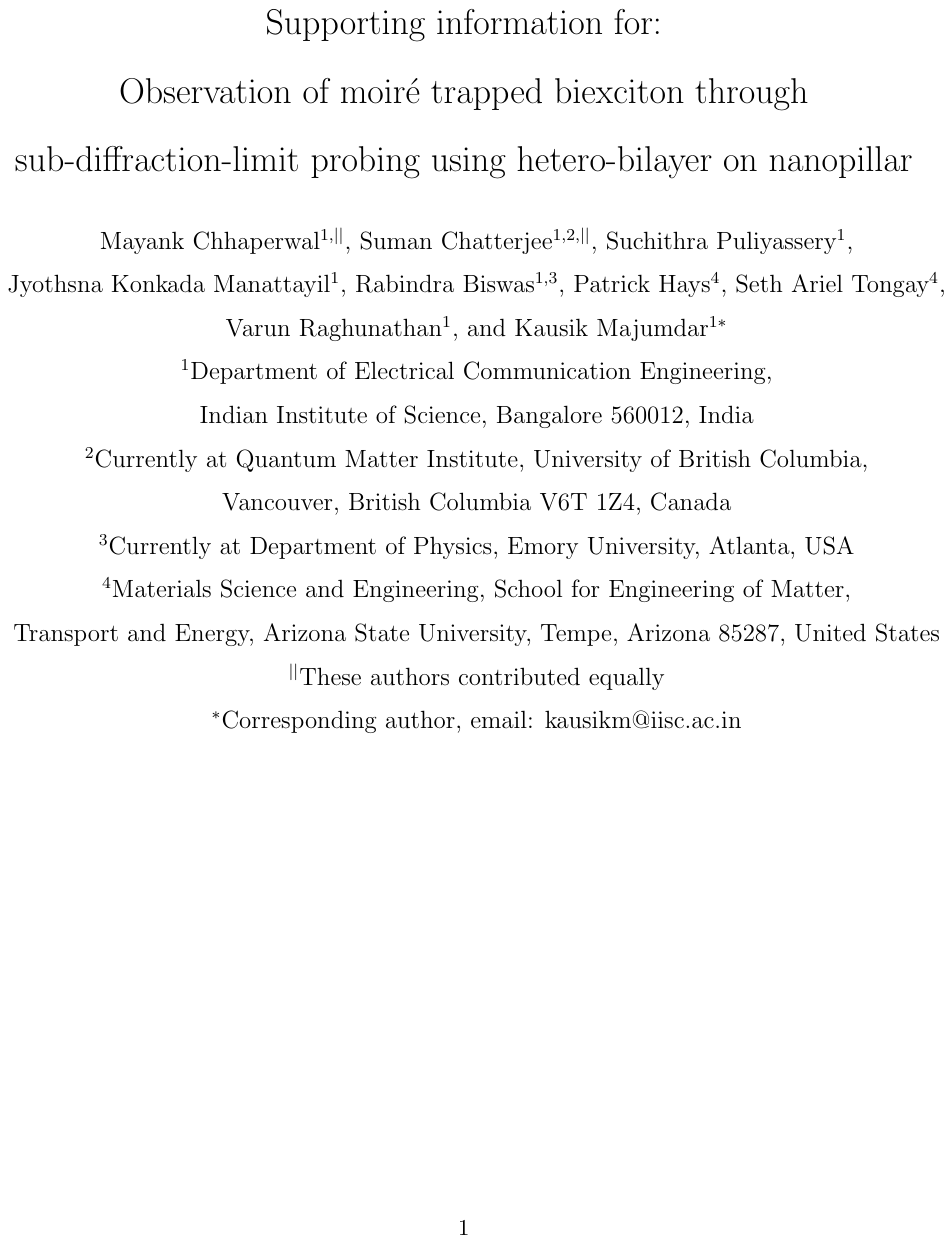}
\end{document} <<